\documentclass{article}

\usepackage{graphicx}

\begin{document}

\title{Separation of the Alfv\'{e}n mode from a turbulent velocity field using wavelets}
\author{Grzegorz Kowal and Alex Lazarian\footnote{Department of Astronomy, University of Wisconsin, 475 North Charter Street, Madison, WI 53706, USA, kowal@astro.wisc.edu}}

\maketitle

\begin{abstract}
We report the first attempt to mode decomposition of the velocity field using wavelet transforms in the application to the magnetized turbulence analysis. We compare results of the separation with respect to the global and local mean magnetic field for models with a strong and weak external magnetic fields. We show, that for models with a strong external magnetic field our results are relatively consistent for both methods. However, for superAlfv\'{e}nic turbulence with spatially variable direction of the local mean magnetic field the differences are substantial.
\end{abstract}

\vspace{0.5in}

Wavelet transform \cite{antoine99} of a one dimensional (1D) signal $s(x)$ produces a function $S(a,b)$ of a dilatation or contraction parameter $a$ and a translation $b$ calculated from the equation:
\begin{equation}
 s(x) \mapsto S(a, b) = a^{-1/2} \int_{-\infty}^{\infty} {\overline{\psi \left( \frac{x - b}{a} \right)} \, s(x) dx} ,
 \label{eq:wt}
\end{equation}
where $\psi$ is an analyzing function called a wavelet. The wavelet transform can be also performed in the Fourier domain
\begin{equation}
 S(a, b) = a^{1/2} \int_{-\infty}^{\infty}{\overline{\hat{\psi}\left(a k\right)} \hat{s}\left(k\right) e^{i b k} dk}.
 \label{eq:wt_fourier}
\end{equation}

In general, the function $\psi$ can be arbitrary. However, for usefulness of the transform, we use functions of particular properties. The function $\psi$ is assumed to be well localized both in space and frequency domains. In addition, $\psi$ must satisfy the admissibility condition:
\begin{equation}
 \int_{-\infty}^{\infty} | \hat{\psi} (k)  |^2 \frac{d k}{|k|} < \infty.
\end{equation}
This condition guarantees the reversibility of the wavelet transform. In most cases this can be reduced to the requirement that $\psi$ has zero mean:
\begin{equation}
 c_{\psi} = \int_{-\infty}^\infty \psi(x) dx = 0.
\end{equation}
In addition, we can require for $\psi$ to have a certain number of vanishing moments:
\begin{equation}
 \int_{-\infty}^\infty {x^n \psi(x) dx} = 0, \ n = 0, 1, ..., N,
\end{equation}
which improves the efficiency of $\psi$ at detecting singularities in the signal.

The transformation $s(x) \mapsto S(a, b)$ may be inverted exactly, which yields to a reconstruction formula:
\begin{equation}
 s(x) = c_{\psi}^{-1} \int_{-\infty}^{\infty}{db \int_0^{\infty}{\frac{da}{a^2} \psi\left( \frac{x - b}{a} \right) S(a,b)}}.
 \label{eq:iwt}
\end{equation}

The mode decomposition is usually considered with relation to simulation data, which in the case of turbulence are more than one dimensional. To apply the wavelet transform, we have to define the transform in the multidimensional form. Following \cite{meneveau91} we can define multidimensional wavelet transform as
\begin{equation}
 S(a, {\bf b}) = a^{-n/2} \int_{-\infty}^{\infty}{\psi\left( \frac{{\bf x} - {\bf b}}{a} \right) s({\bf x}) d^n{\bf x}},
\end{equation}
where ${\bf x}$ and ${\bf b}$ are $n$-dimensional vectors of position and translation, respectively. We assume here an isotropic wavelet $\psi({\bf x}) = \psi(|{\bf x}|)$. The transform in Fourier domain and inverse transform can be expressed in analogous way from Eqs.~(\ref{eq:wt_fourier}) and (\ref{eq:iwt}).

The function $S(a, {\bf b})$ can be interpreted here as an amplitude or energy contained by a wave packet of dilated or contracted by parameter $a$ at position ${\bf b}$. This interpretation allows to treat a vector field like a set of three scalar fields, so its wavelet transform is simplified to three transforms of each field component.

We present results obtained for two 3D numerical experiments of compressible magnetohydrodynamic (MHD) turbulence for the Alfv\'{e}nic Mach numbers ${\cal M}_A$ about $0.7$ ($B_0=1.0$) and $\sim 7$ ($B_0=0.1$). $B_0$ is the initial external magnetic field. Both models have the same initial pressure which corresponds to the sonic Mach number equal to about $0.7$. We understand the Mach number defined as a mean value of the ratio of the absolute value of the local velocity $V$ to the local value of the characteristic speed $a$ or $V_A$ (for the sonic and Alfv\'{e}nic Mach numbers, respectively). Both models are calculated with resolution $256^3$.

\begin{figure}
 \centering
 \includegraphics[width=1.5in]{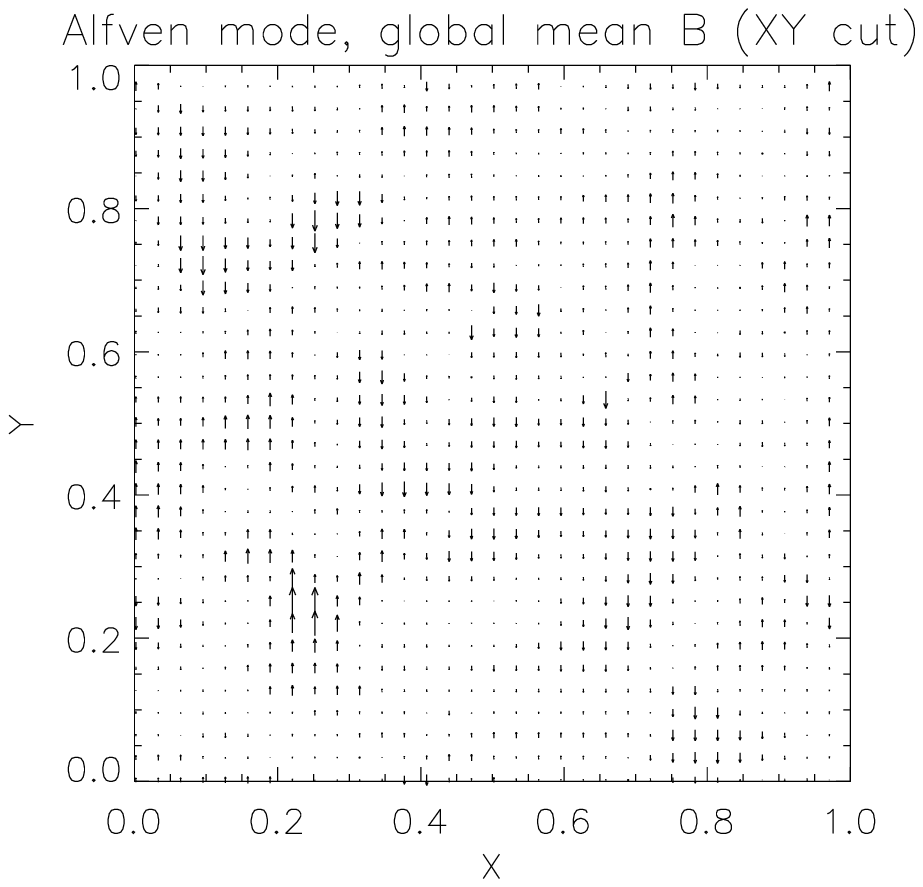}
 \includegraphics[width=1.5in]{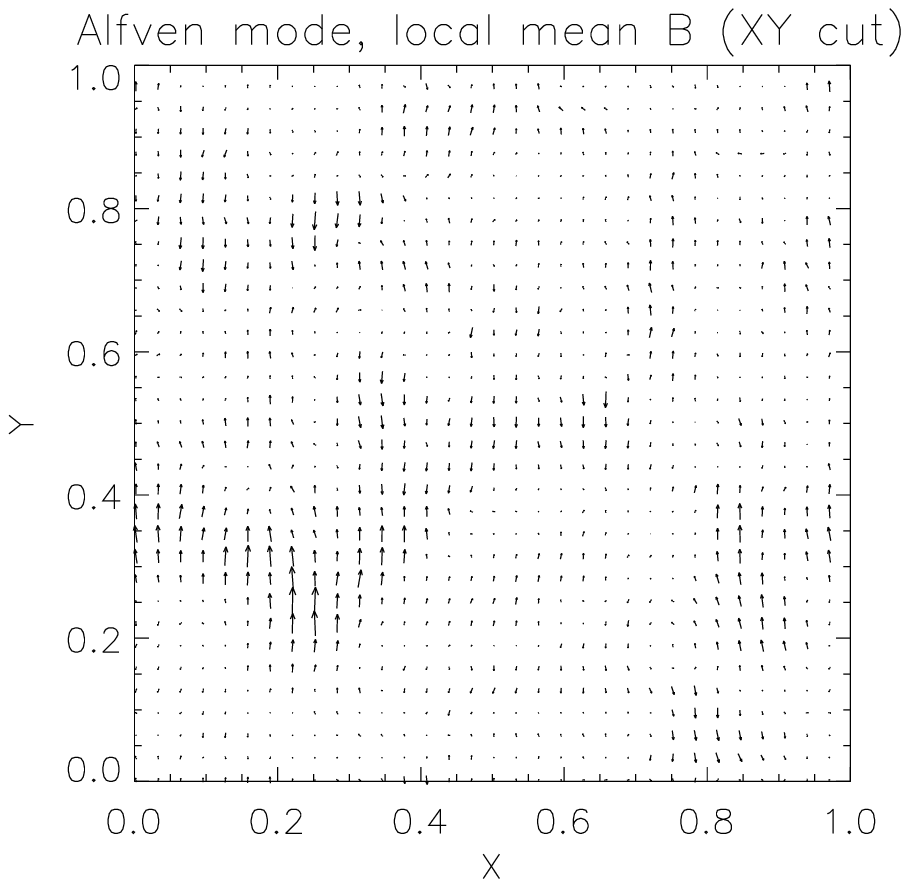}
 \includegraphics[width=1.5in]{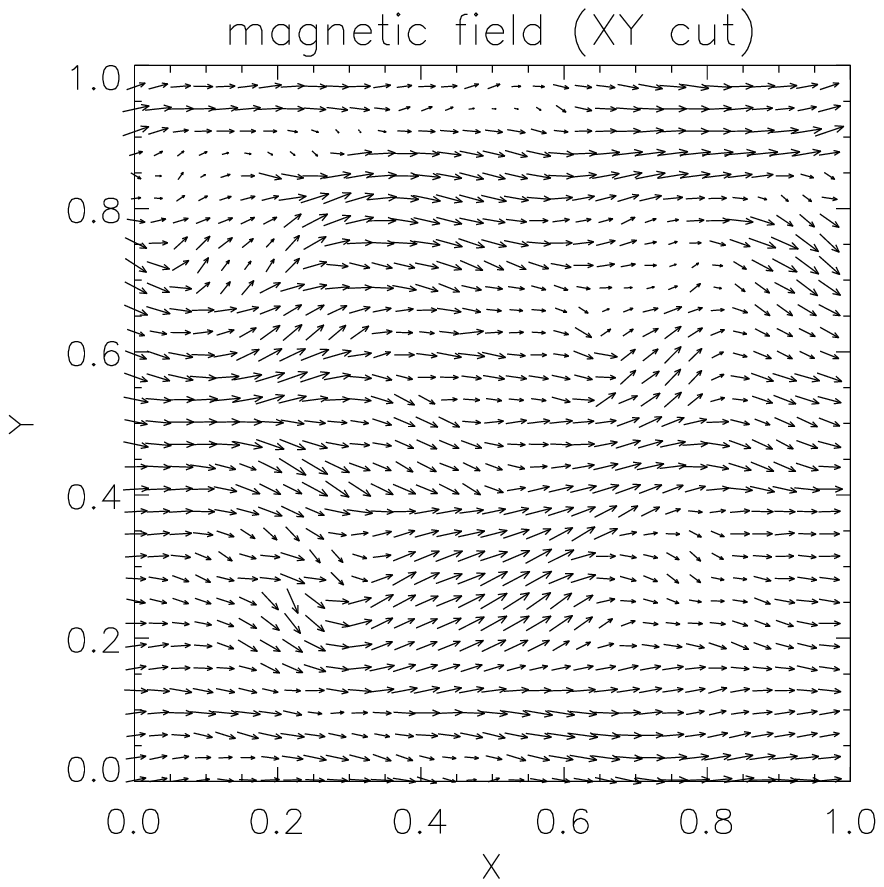}
 \includegraphics[width=1.5in]{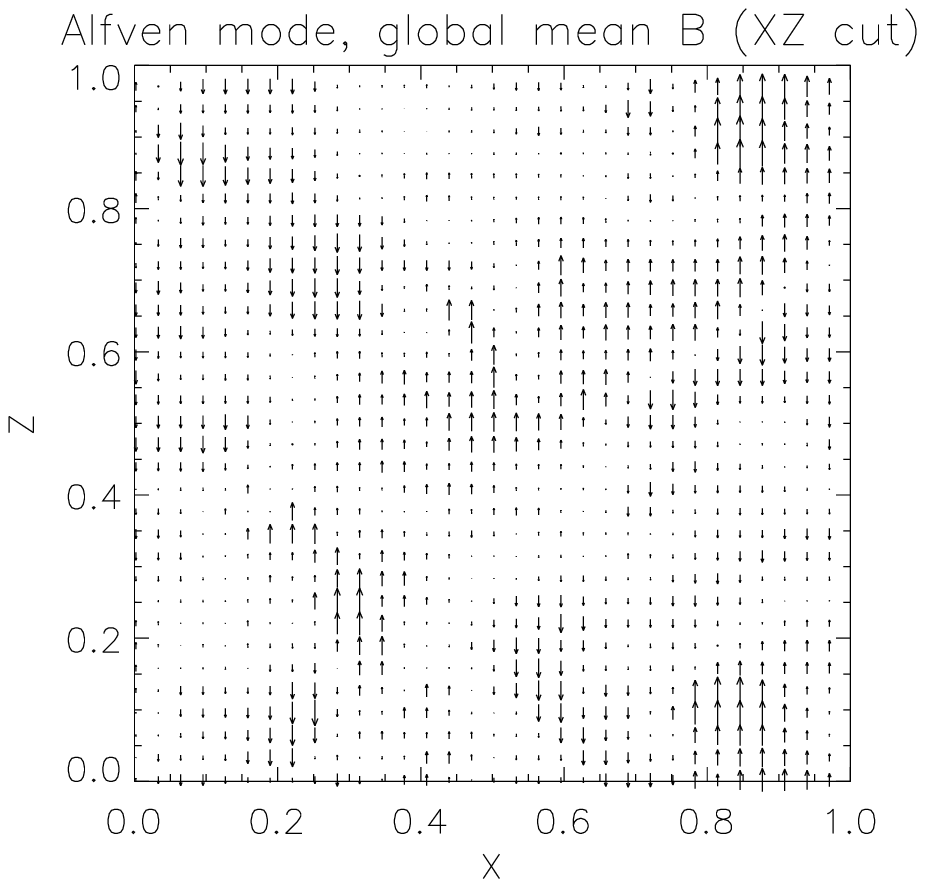}
 \includegraphics[width=1.5in]{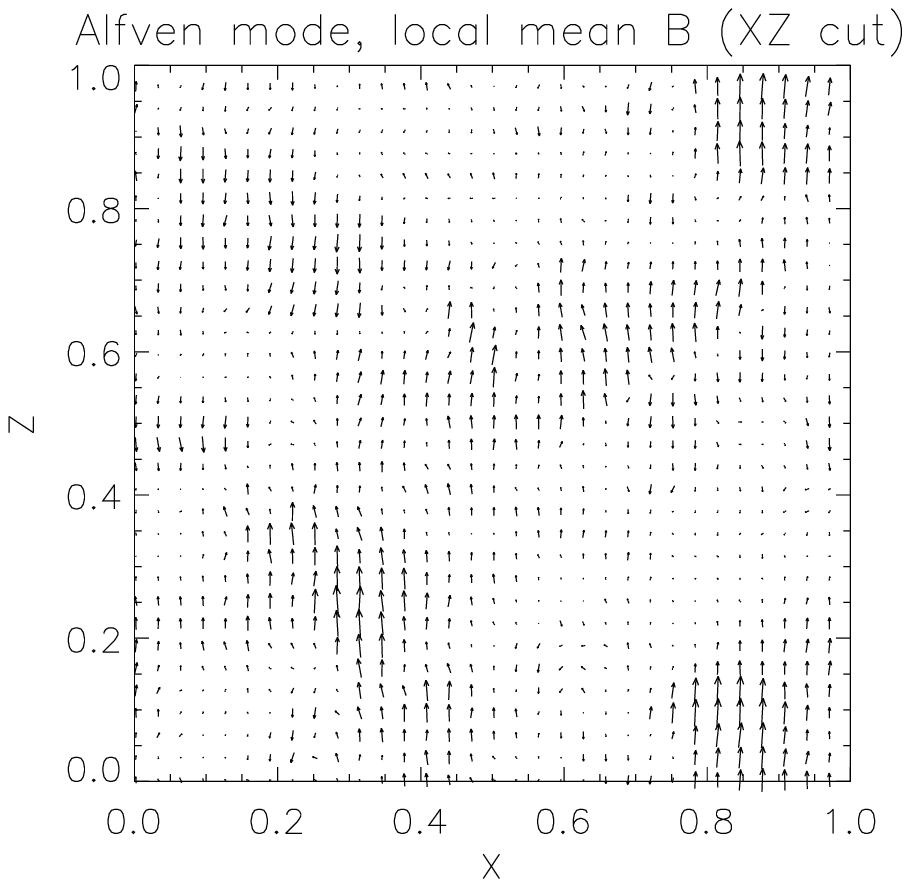}
 \includegraphics[width=1.5in]{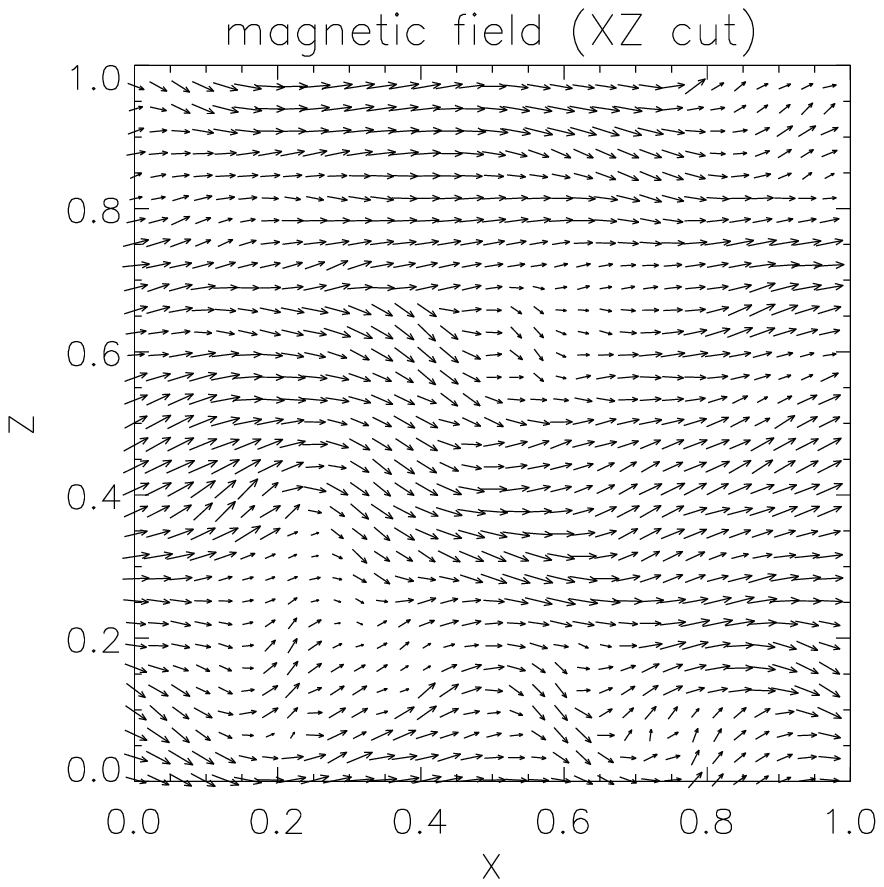}
 \includegraphics[width=1.5in]{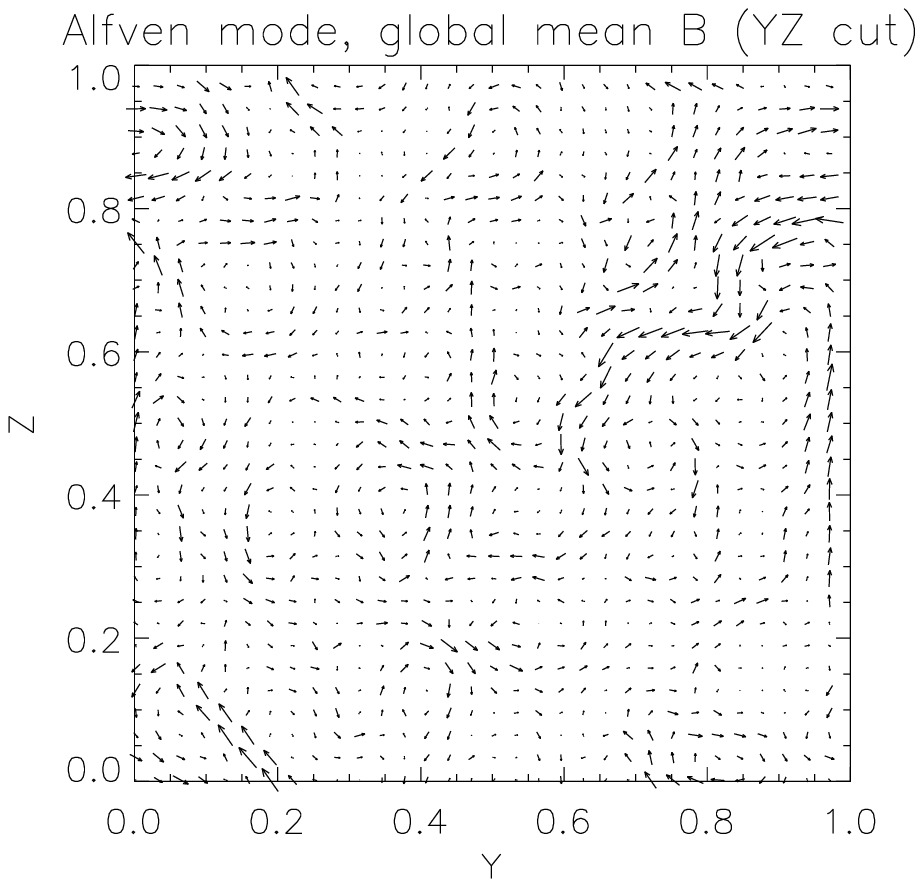}
 \includegraphics[width=1.5in]{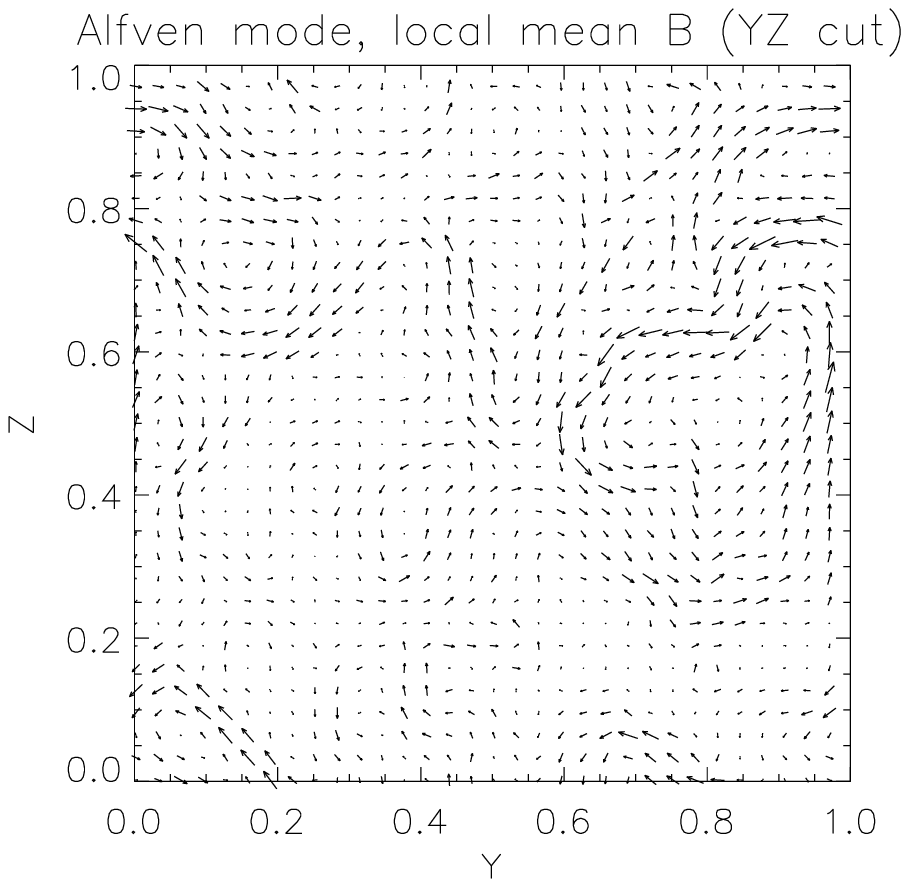}
 \includegraphics[width=1.5in]{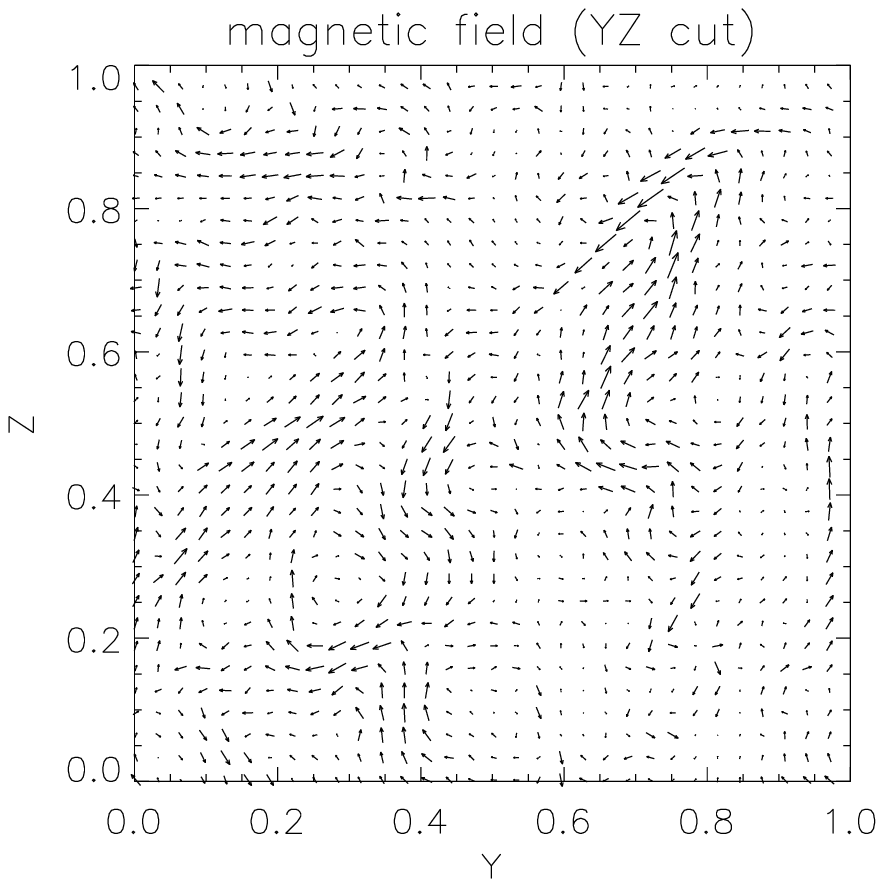}
 \caption{Alfv\'{e}n mode for the subAlfv\'{e}nic turbulence. We show vectors along three intersections of the model domain: XY, XZ and YZ (top, middle and bottom rows, respectively). Left column shows the Alfv\'{e}n mode separated with the respect to the global mean magnetic field. Middle column shows exactly the same intersections, but for the Alfv\'{e}n mode separated with respect to the local mean magnetic field. For comparison we also show the magnetic field on the right column. \label{fig:sub_alfven}}
\end{figure}

We start from the model with a strong magnetic field $B_0=1$. We expect that both decompositions, with respect to the global and local mean magnetic field, should give approximately the same results. The Alfv\'{e}n mode is perpendicular to the mean magnetic field, so its X-component should be significantly smaller, because ${\bf B}_0$ has only X-component different from zero. We present corresponding vectors of the field at three intersections (XY, XZ and YZ) in Figure \ref{fig:sub_alfven}. We notice, that our expectations were correct. In the top and middle plots on the left column we see, that X-component of the field is zero everywhere. These plots are for the decomposition with respect to the direction of the mean magnetic field taken globally. What should we expect if we take into account the direction of the local mean magnetic field? In this case, we expect results very similar. Indeed, looking in the top and middle plots on the middle column of Figure~\ref{fig:sub_alfven} we see that the X-component of the Alfv\'{e}n mode is marginal. Total energies of the particular components for decomposition with the respect to the global mean magnetic field are: 0.0, 0.06, 0.07 for X, Y and Z-component, respectively. For decomposition with respect to the local mean magnetic field these energies are: 0.002, 0.1, 0.09. We see that in the latter case, the energy of X-component is about two orders of magnitude smaller then for the other two components.

The agreement in the strength of individual components, however, is not enough to prove, that both methods are consistent. We should expect some similarities in the pattern of the fields. Again, we compare plots for corresponding intersections for both types of decomposition. We see, that the patterns are not reconstructed perfectly. However, the most of the structure is similar. If we include in comparison the plots from the right column of Figure~\ref{fig:sub_alfven} presenting the magnetic field for the same intersections, we see, that those parts of the fields which are similar in plots of the left and middle column correspond to the places where the magnetic field is strong and approximately directed along X-direction. For YZ-intersection, which shows a plane perpendicular to the global mean magnetic field, the correspondence of fields obtained by two methods is in very good agreement.

\begin{figure}
 \centering
 \includegraphics[width=1.5in]{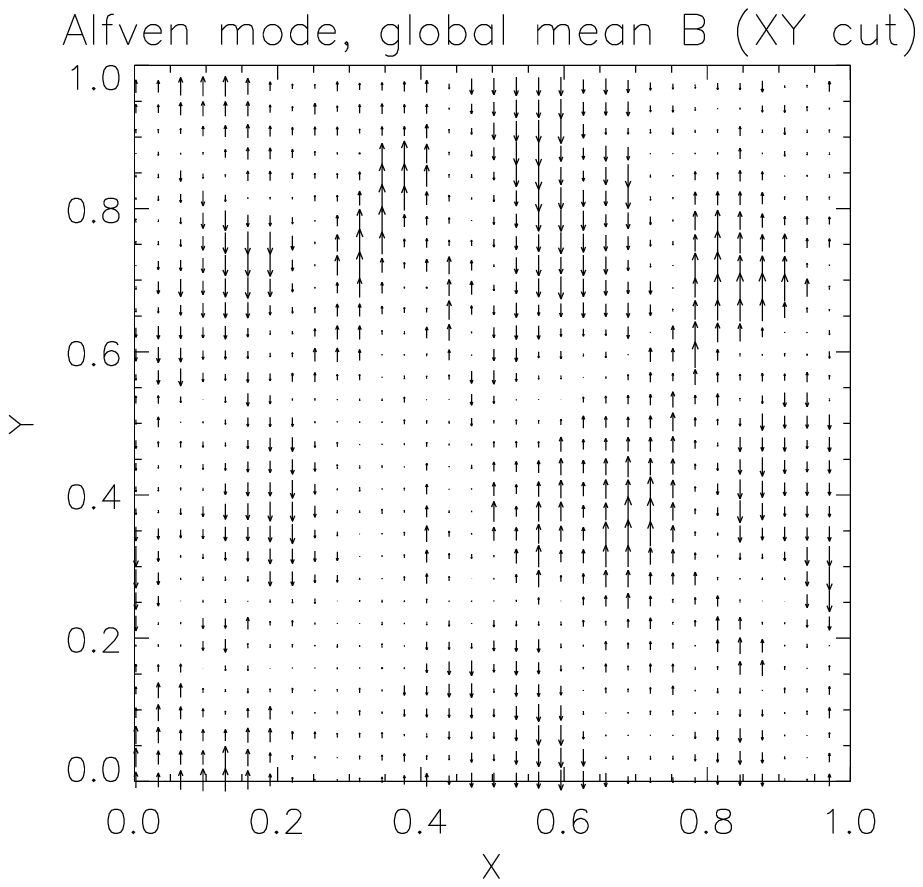}
 \includegraphics[width=1.5in]{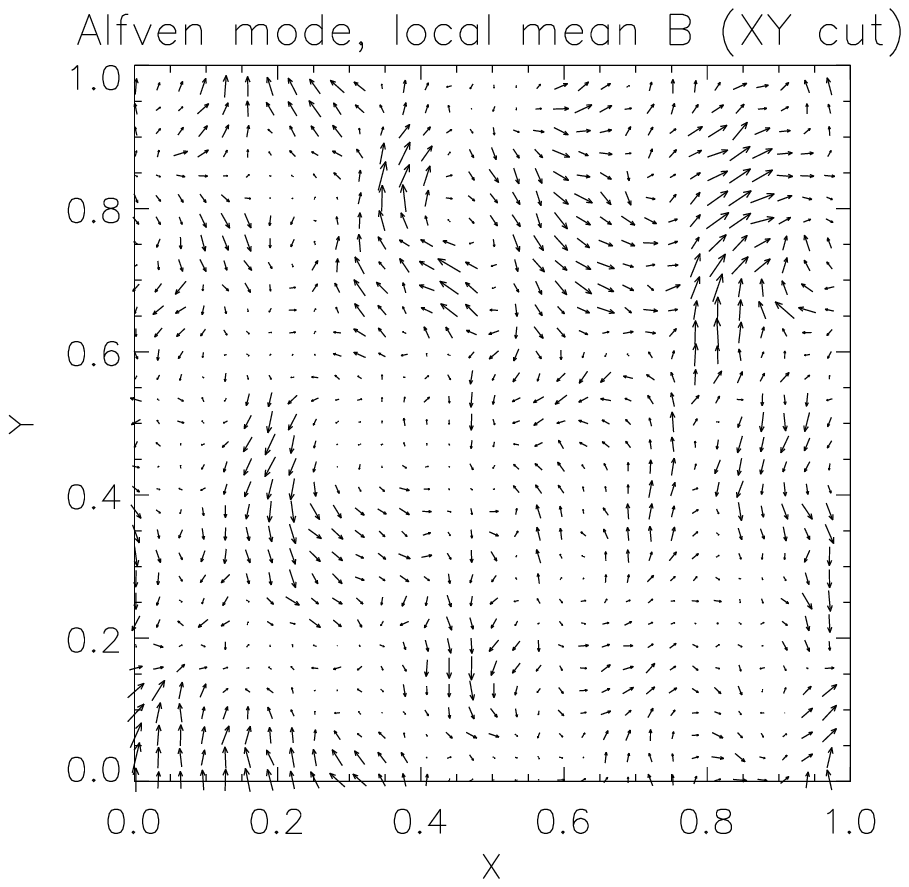}
 \includegraphics[width=1.5in]{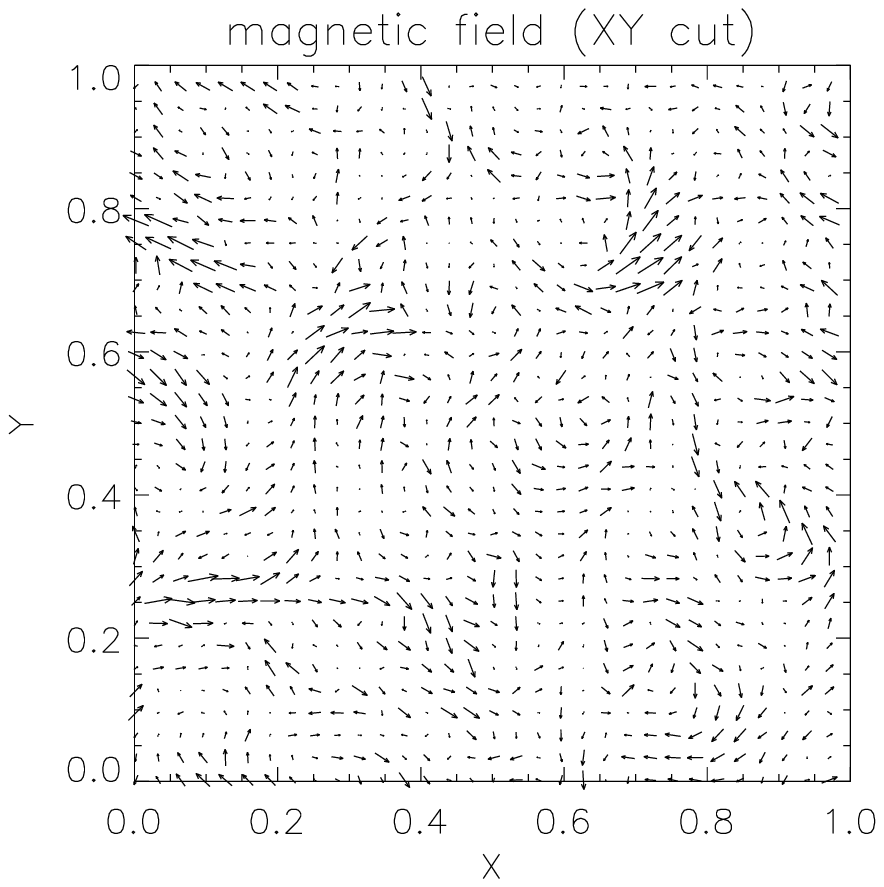}
 \includegraphics[width=1.5in]{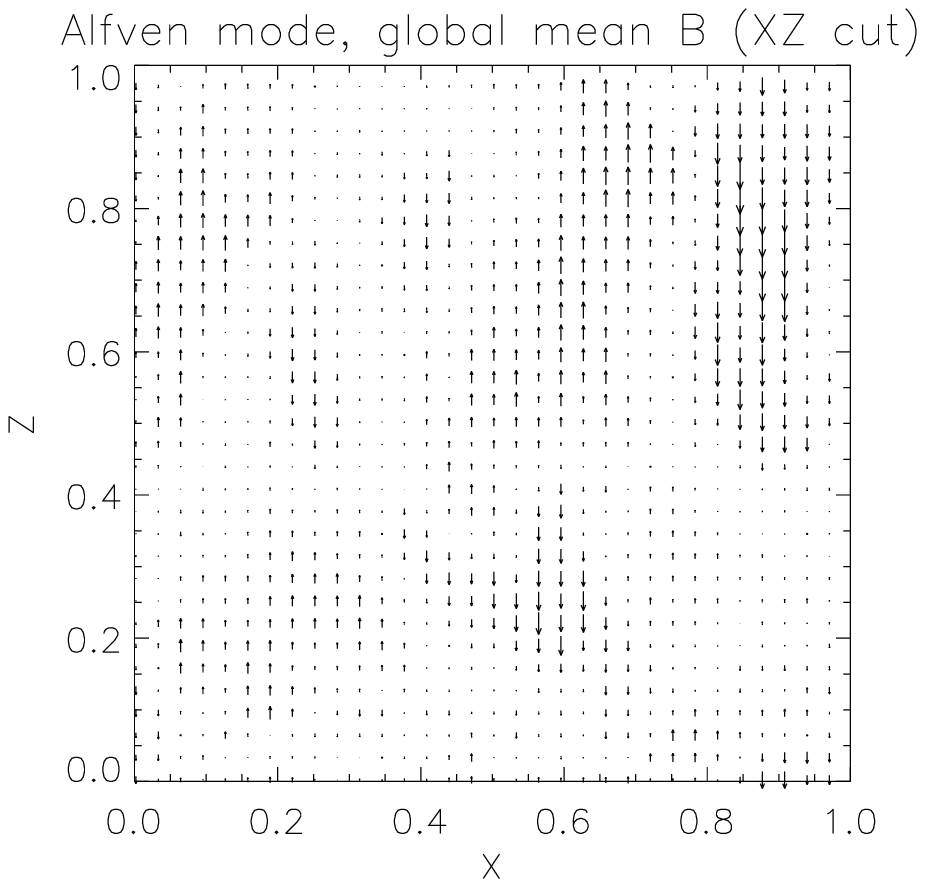}
 \includegraphics[width=1.5in]{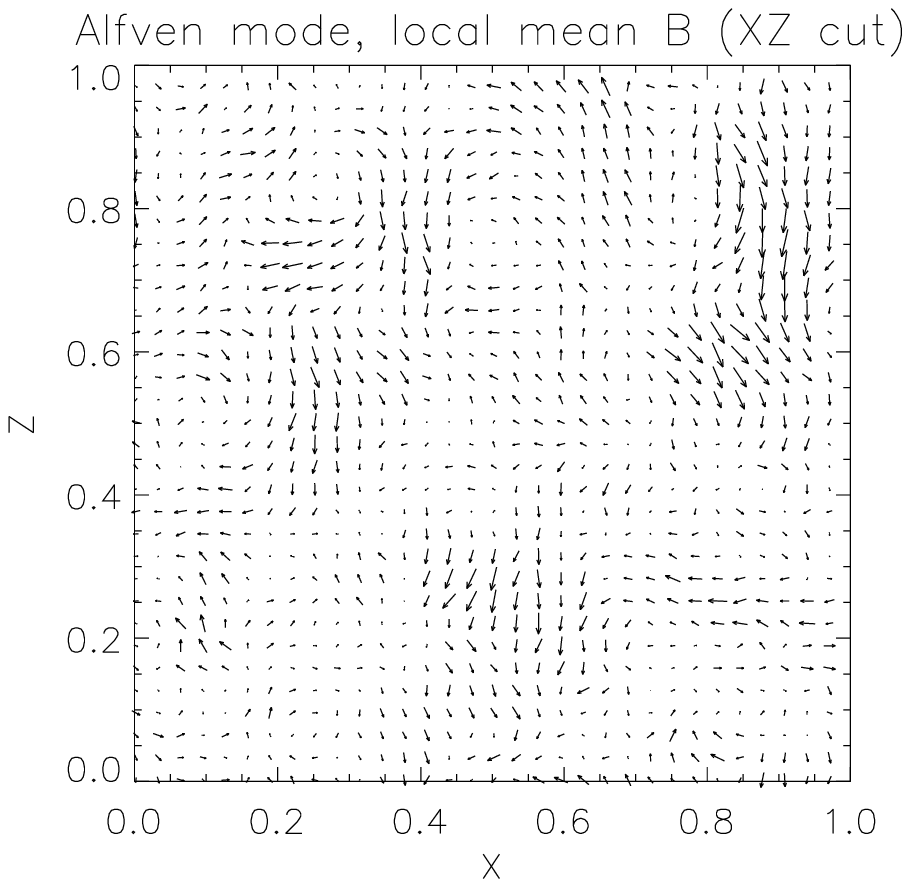}
 \includegraphics[width=1.5in]{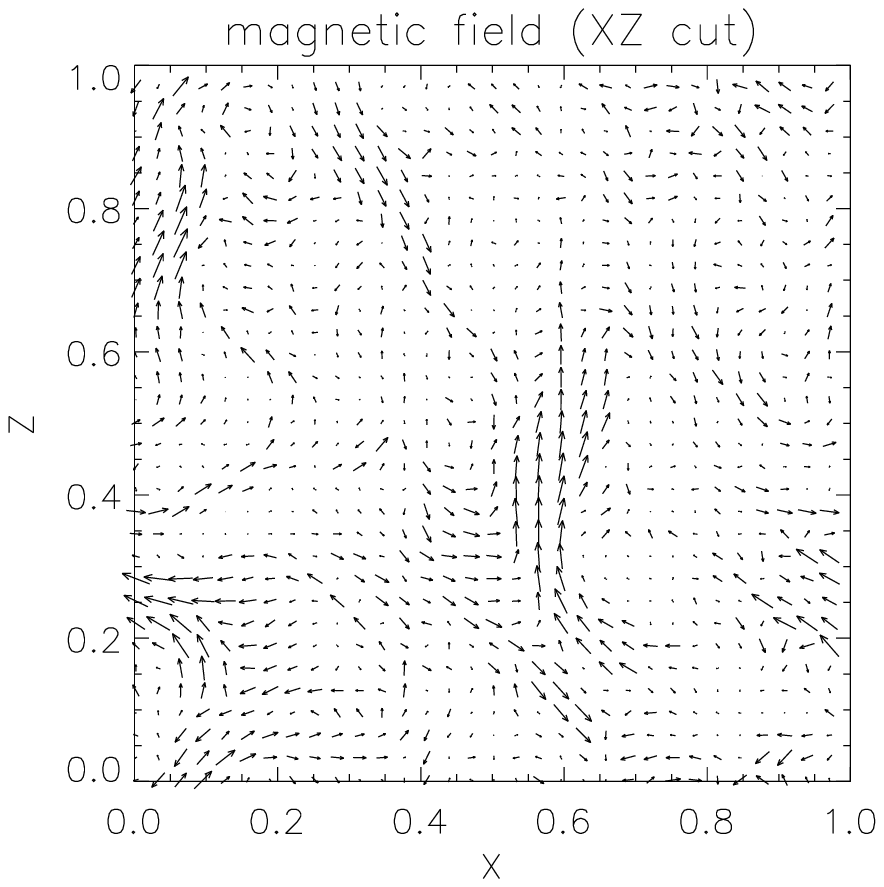}
 \includegraphics[width=1.5in]{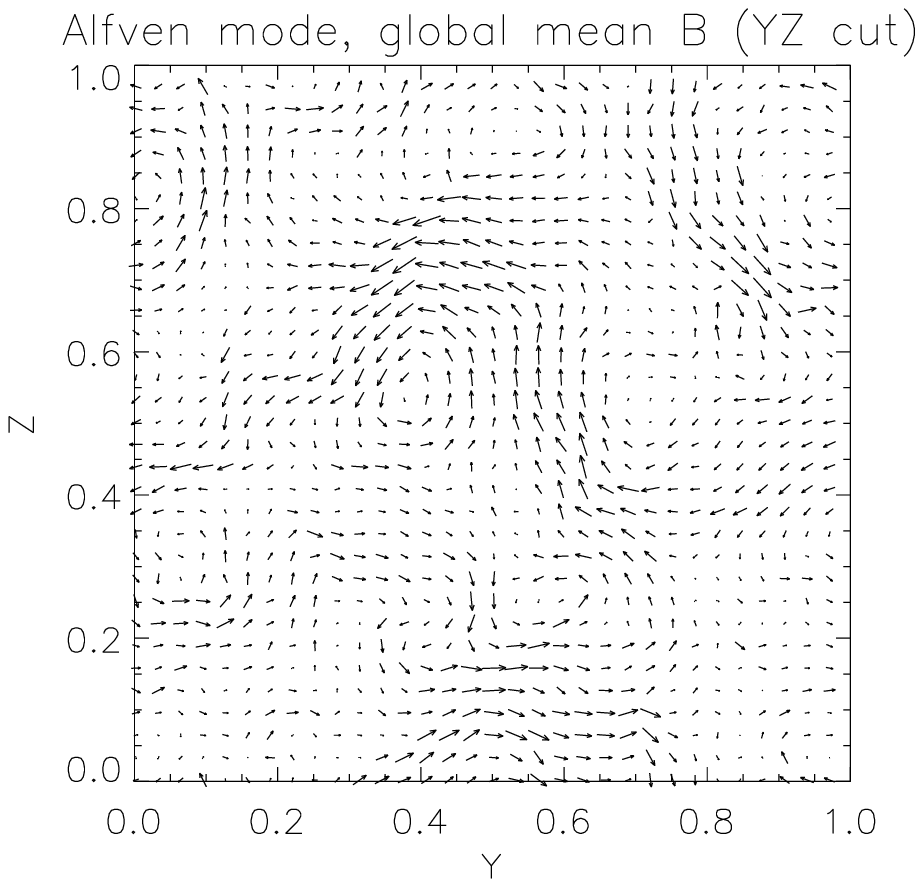}
 \includegraphics[width=1.5in]{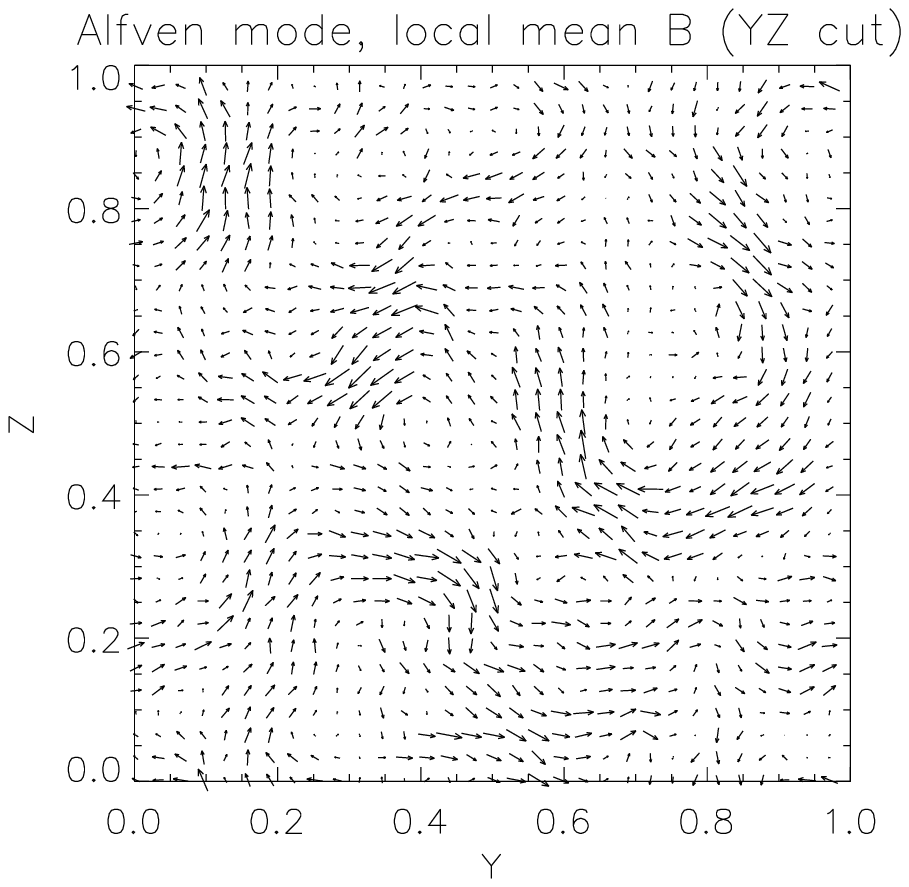}
 \includegraphics[width=1.5in]{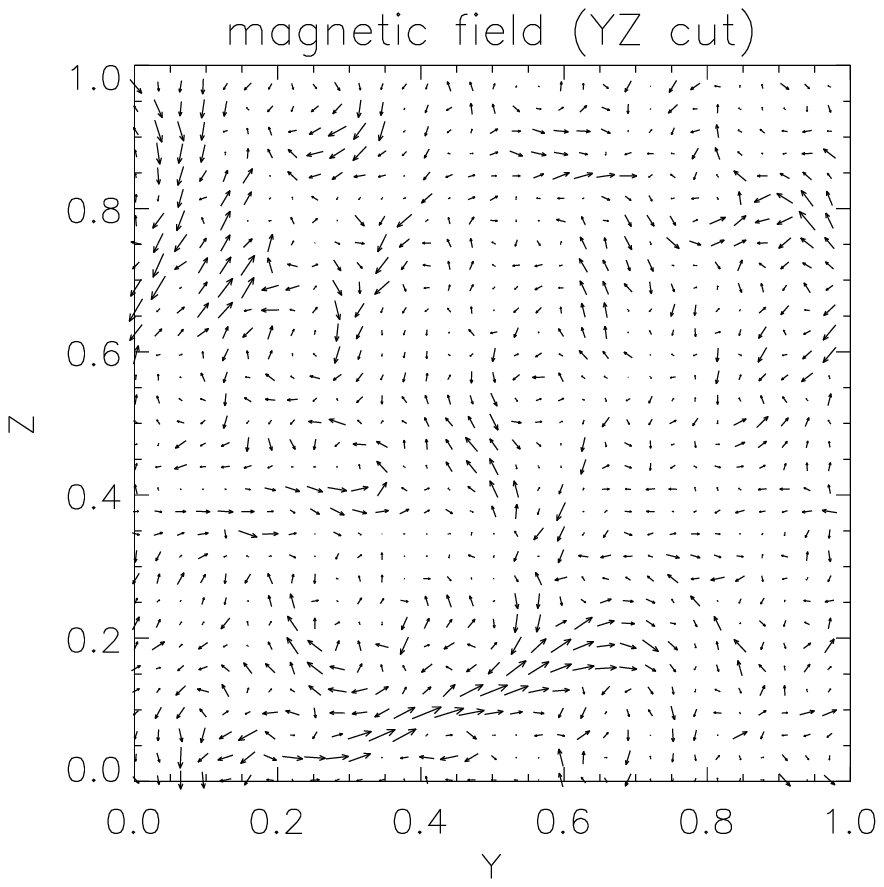}
 \caption{Alfv\'{e}n mode for the superAlfv\'{e}nic turbulence. We show vectors along three intersections of the model domain: XY, XZ and YZ (top, middle and bottom, respectively). Left column shows the Alfv\'{e}n mode separated with the respect to the global mean magnetic field. Middle column shows exactly the same intersections, but for the Alfv\'{e}n mode separated with respect to the local mean magnetic field. For comparison we also show the magnetic field on the right column. \label{fig:sup_alfven}}
\end{figure}

We showed that in the case of turbulence simulations with a strong magnetic field both methods give quite similar separation of the Alfv\'{e}n mode. This signifies, that the assumption of the direction of the global mean magnetic field in decomposition can be applied for velocity field separation. However, could it be still justified in the case of turbulence with a weak mean magnetic field? The answer seems to be 'no'. Figure~\ref{fig:sup_alfven} shows the comparison of Alfv\'{e}n modes obtained with the help of both methods from the superAlfv\'{e}nic turbulence. The comparison shows clearly large differences in the field patterns between the plots on the left and middle columns. If we include the local mean magnetic field in the separation procedure, we will obtain the X-component of strength comparable to the other two components: Y and Z. For separation with respect to the global mean magnetic field we get energies of X, Y and Z-components equal to about 0.0, 0.1, 0.1. After taking into account the local mean magnetic field, these energies are 0.04, 0.1, 0.08, respectively.

Comparison of plots  of the Alfv\'{e}n mode separated with respect to the local mean magnetic field to plots of magnetic field reveals a tendency, which should be expected. Precisely, the vectors of the Alfv\'{e}n part of the velocity field are perpendicular to the vectors of magnetic field at the same places. This is additional confirmation for the validity of the method based on the wavelet transform.

We have presented the separation of the velocity field into the Alfv\'{e}n mode based on the wavelet transform in the application to the magnetized turbulence. We compared the validity of this method for models with a strong and weak external magnetic field. We showed that the separation with respect to the global mean magnetic field can be still valid when the external magnetic field is significantly larger the the strength of its fluctuations. However, in the case of violating this condition, this assumption cannot be valid anymore. In this case we are obligate to use methods based on wavelet transforms.


\end{document}